\documentclass[twocolumn,showpacs,preprintnumbers,amsmath,amssymb]{revtex4}

\usepackage{graphicx}
\usepackage{dcolumn}
\usepackage{bm}

\begin{document}

\preprint{APS/123-QED}

\title{Single electron control in n-type semiconductor quantum dots using non-Abelian holonomies
generated by spin orbit coupling}
\author{S.-R. Eric Yang$^{1,2}$\footnote{ eyang@venus.korea.ac.kr}}
\author{N.Y. Hwang$^{1}$}
\affiliation{$^{1}$Physics Department, Korea  University, Seoul Korea \\
$^{2}$School of Physics, Korea Institute for Advanced Study, Seoul Korea}


\begin{abstract}
We propose  that   n-type semiconductor quantum dots with the Rashba and Dresselhaus spin orbit interactions 
may be used for   single electron manipulation through adiabatic transformations between degenerate states.
All the energy levels are  discrete in quantum dots and possess  a double degeneracy due to time reversal symmetry
in the presence of the Rashba and/or Dresselhaus spin orbit coupling terms.
We find that  the presence of double degeneracy 
does not necessarily give rise to  a finite non-Abelian (matrix) Berry phase.
We show  that a  distorted   two-dimensional harmonic potential 
may give rise to non-Abelian Berry phases.
The presence of the non-Abelian Berry phase may  be tested experimentally 
by measuring  the optical dipole transitions.
\end{abstract}

\pacs{71.55.Eq, 71.70.Ej, 03.67.Lx, 03.67.Pp}
\maketitle

\begin{figure}[hbt]
\begin{center}
\includegraphics[width = 0.15 \textwidth]{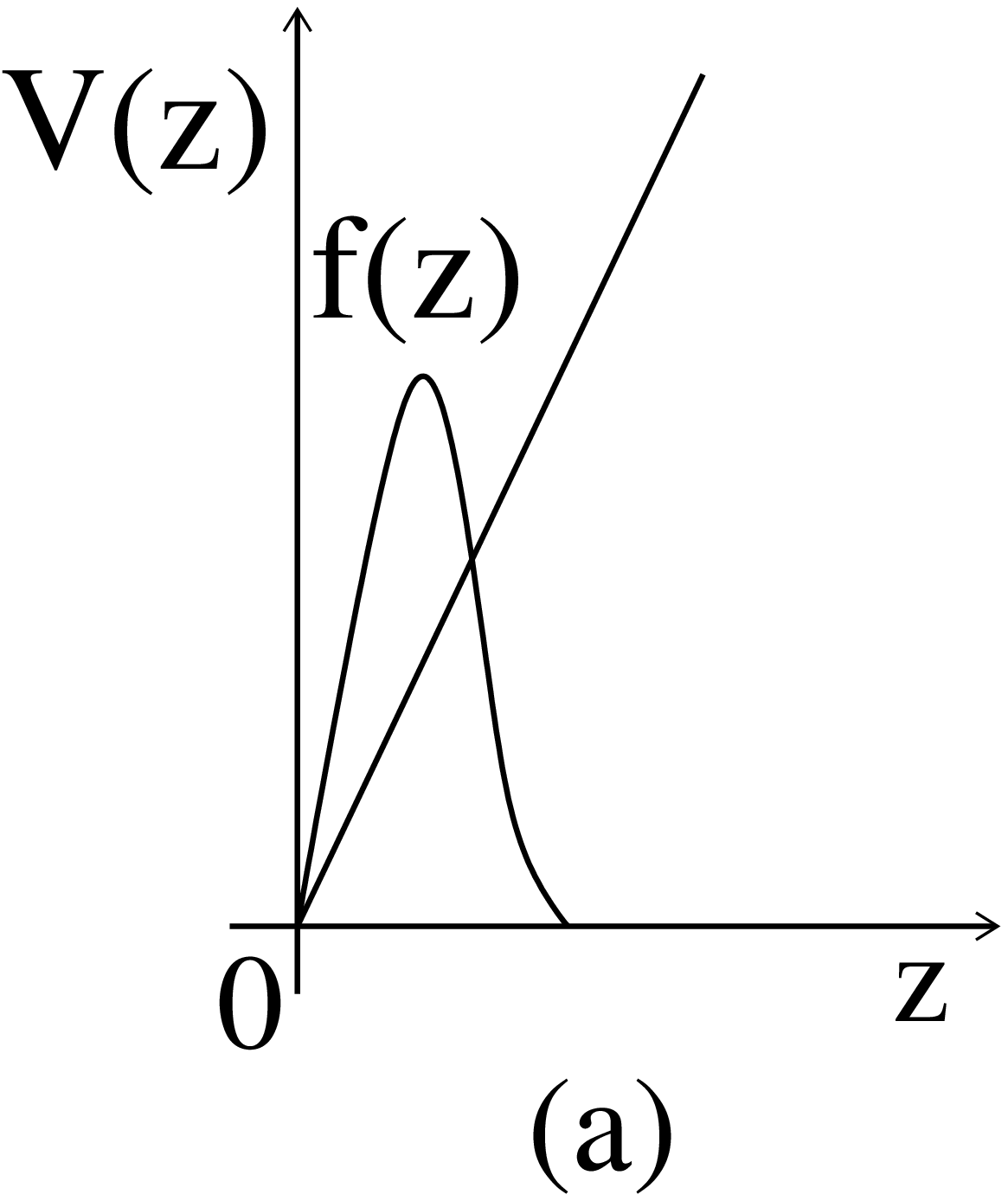}
\includegraphics[width = 0.15 \textwidth]{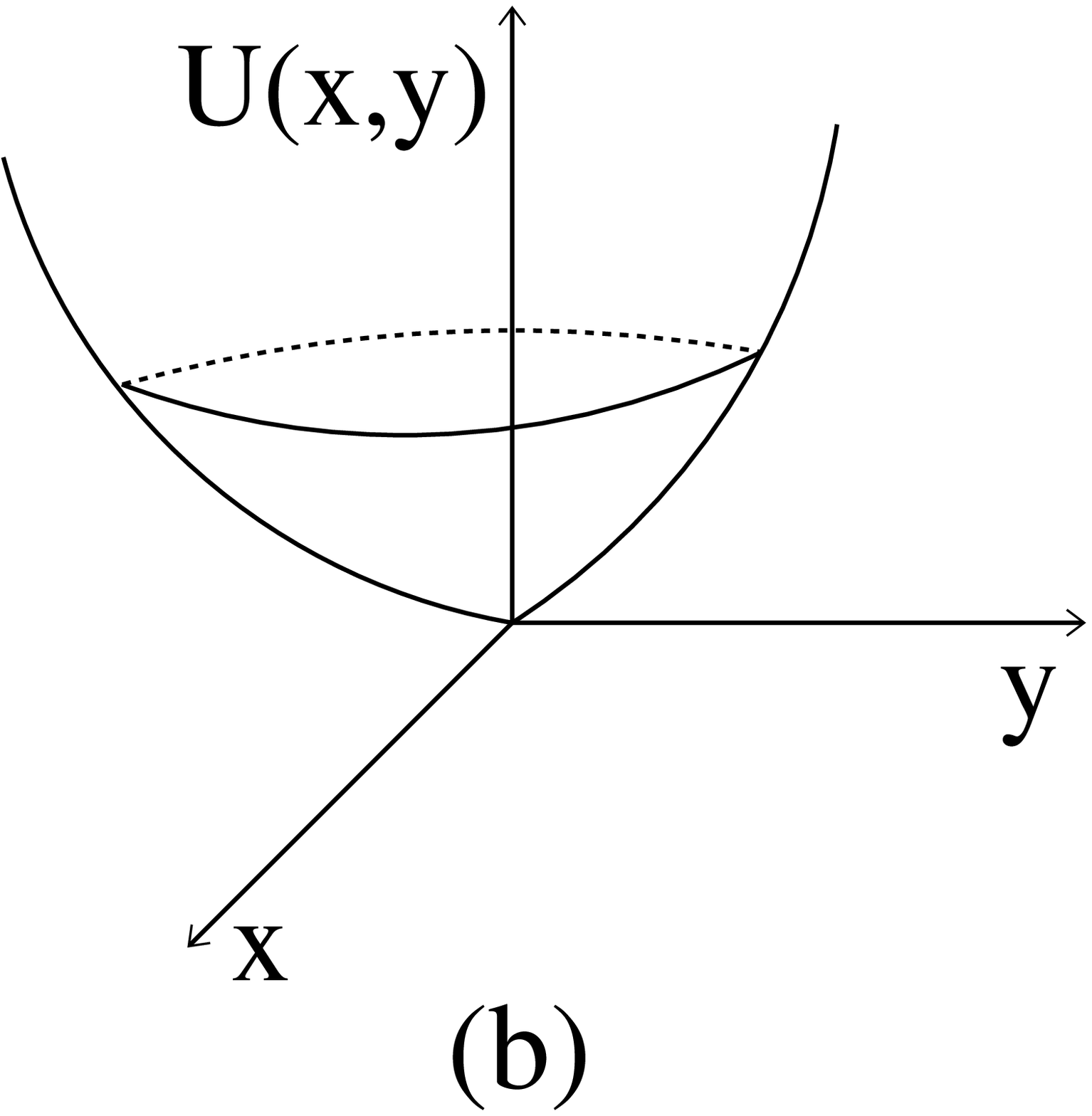}
\includegraphics[width = 0.15 \textwidth]{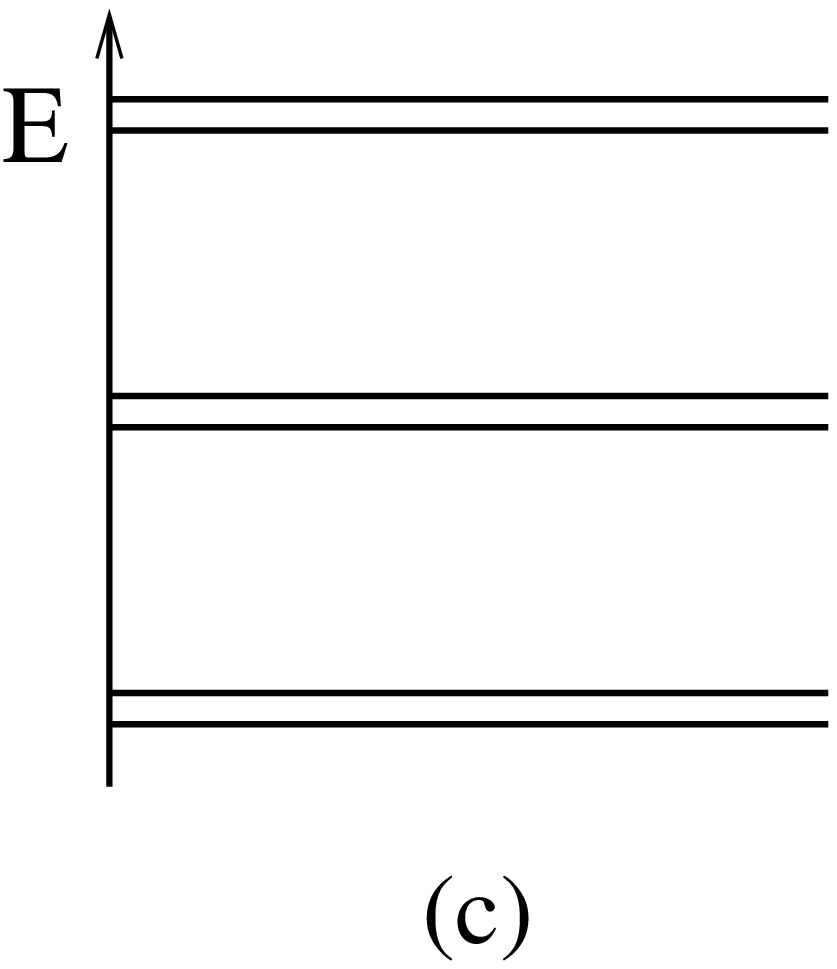}
\includegraphics[width = 0.3 \textwidth]{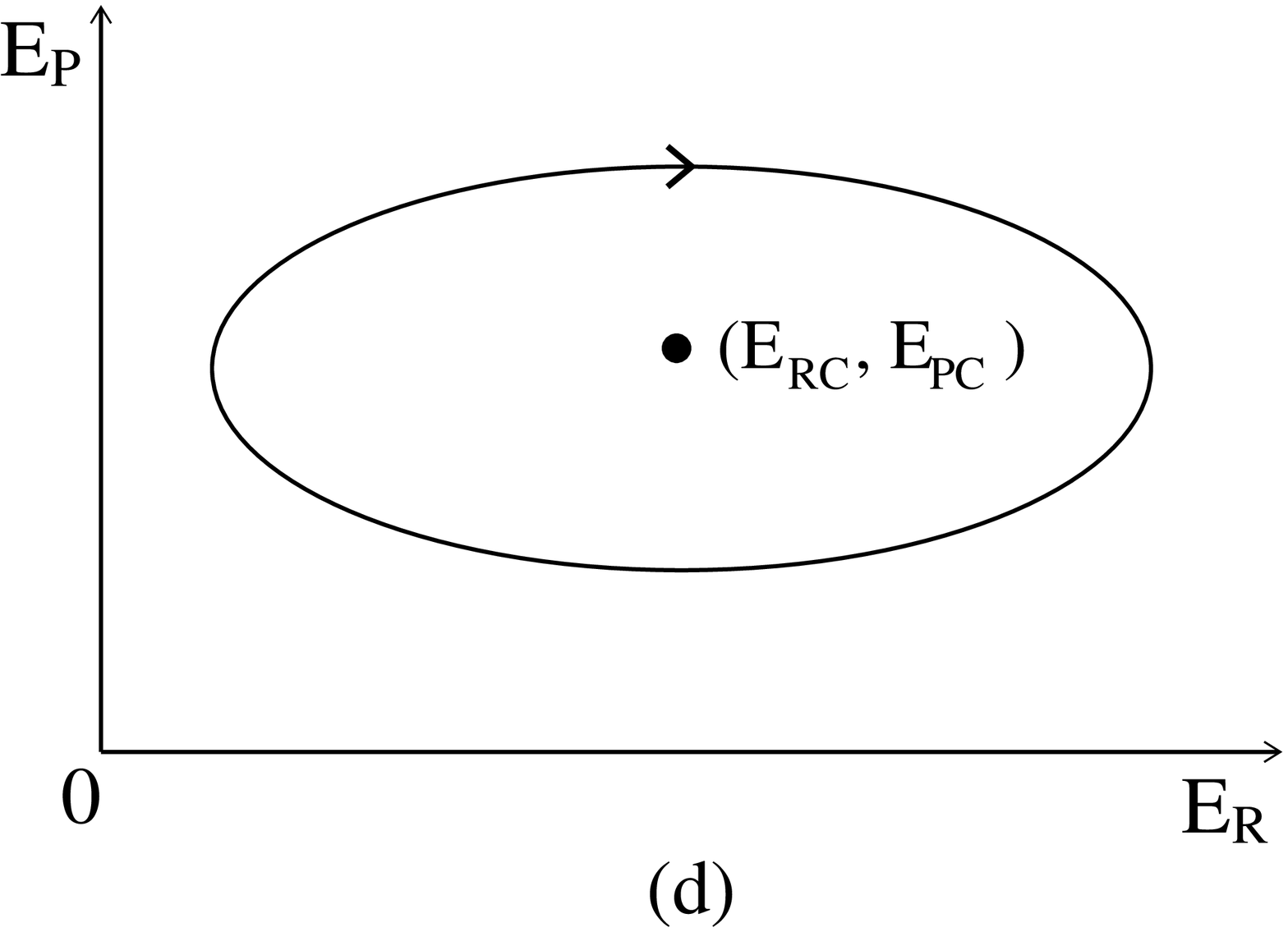}
\caption{(a) An electric field along the z-axis quantizes the electronic motion in a triangular potential along the axis.
In our work  the triangular potential is sufficiently strong and  only the lowest energy subband is included.
The structural inversion asymmetry in $V(z)$ leads to the Rashba interaction.
An adiabatic change can be induced by changing the magnitude of the electric field.
(b) The two-dimensional electronic motion is quantized in a distorted harmonic potentials  
$U(x,y)=\frac{1}{2}m^*\omega^2_x x^2+\frac{1}{2}m^*\omega^2_y y^2+\epsilon'y$.
An adiabatic change can be induced by changing the magnitude of the distortion  potential.
(c) In the presence of the Rashba and/or Dresselhaus spin orbit coupling terms 
each discrete eigenstate  of a semiconductor quantum dot has 
a double degeneracy due to time reversal symmetry in the absence of a magnetic field.  The three
lowest energy levels are shown schematically (Each pair is doubly degenerate).
(d) A schematic drawing of a cyclic adiabatic path. The parameters   $E_R$ and $E_p$
depend on   the magnitudes of the electric field and the deviation from the two-dimensional harmonic potentials.}
\label{fig:setup}
\end{center}
\end{figure}

\section{Introduction}

Single electron control in semiconductor quantum dots  would be valuable for spintronics, quantum information, and spin qubits\cite{Aws}.
Adiabatic time evolution of degenerate eigenstates of a quantum system provides a means
for controlling individual quantum states\cite{Sha,Wil,Zee}. 
They exhibit  non-Abelian gauge structures
and often  give finite  non-Abelian Berry phases (they are also called matrix Berry phase  or holonomic phase).
These phases often depend on the geometry of the path traversed in the parameter space of the Hamiltonian.
Nuclei\cite{jones2}, superconducting nanocircuits\cite{fal}, optical systems\cite{pach2}, and 
atomic systems\cite{Mead,duan,Un} 
have such degenerate quantum states.
It has been shown that universal quantum computation is possible by means of non-Abelian unitary
operations\cite{Zan,Pach}.
Manipulation is  expected to be stable
since symmetries of the Hamiltonian that give rise to degeneracy are not broken
during the adiabatic transformations.  
However, the degree of its stability is under investigation\cite{Sol1}.

Recently several interesting possibilities for electronic manipulation in
semiconductors have been proposed.
Spin manipulation of quasi-two-dimensional electrons by time-dependent gate voltage is possible
through the Dresselhaus and Rashba spin-orbit coupling mechanisms \cite{Rashba1}.
A geometric spin manipulation technique based on acceptor states in p-type semiconductors with spin-orbit coupling
was  proposed\cite{Bern}. 
Spin-orbit coupling and a revolving external electric field
may generate holonomic qubit operations  in CdSe nanocrystals \cite{Sere2}.
Holonomic quantum computation using excitons in semiconductor nanostructures has been also proposed\cite{Sol2}.

A matrix Berry  phase is experimentally interesting because it represents mixing between degenerate levels.
Let us explain briefly the basic ideas using a
simple  system possessing  Kramers' double degeneracy\cite{Mead}.  The Hamiltonian depends on some
external parameters $\lambda_p$.
If the system is  in a superposition  state $|\psi'(0)\rangle =c_1(0)|\psi_1(0)\rangle +c_2(0)|\psi_2(0)\rangle$ 
at time t=0  an adiabatic  evolution  of the parameters
$\lambda_p$ can  transform this state into another state $|\psi'(t)\rangle =c_1(t)|\psi_1(t)\rangle +c_2(t)|\psi_2(t)\rangle$ 
after some time $t$. Here the orthonormal basis states $|\psi_i(t)\rangle$ are the instantaneous eigenstates of the Hamiltonian. 
For a cyclic change with the  period  $T$, represented by a closed contour $C$ in the parameter space,
the states $|\psi_1(T)\rangle$ and  $|\psi_2(T)\rangle$ return to the initial states $|\psi_1(0)\rangle$ and 
$|\psi_2(0)\rangle$, but the coefficients
$c_1(T)$ and $c_2(T)$ may not return the initial values.  In such a case a $2\times2$ matrix Berry phase
(non-Abelian Berry phase) $\Phi_C$ is generated
\begin{eqnarray}
\left(
\begin{array}{c}
c_1(T) \\
c_2(T)
\end{array}
\right)=\Phi_C
\left(
\begin{array}{c}
c_1(0) \\
c_2(0)
\end{array}\right).
\end{eqnarray}
This   non-Abelian geometric phase (holonomy)
is connected to non-Abelian gauge potentials, as we discuss below.
During the adiabatic cycle  the degenerate energy  $E(t)$ varies with time. But   hereafter we will 
set $E(t)=0$ since  it can be easily restored when needed.
The expansion coefficients satisfy the time-dependent Schr$\ddot{o}$dinger equation
\begin{eqnarray}
i \hbar \dot{c}_i=-\sum_j A_{i j} c_j \qquad i=1,2.
\label{eq:time_Schrod}
\end{eqnarray}
The matrix elements $A_{ij}$ are given by 
$A_{ij}= \hbar \sum_p(A_p)_{i,j}\frac{d\lambda_p}{dt}$, where the 
sum over $p$ in $A_{ij}$ is meant to be the sum  over $\lambda_p$.
$A_p$ are  $2\times2$ matrices and are called  the non-Abelian vector potentials.
They  are given by
\begin{eqnarray}
(A_p)_{i,j}=
i \langle\psi_i|\frac{\partial\psi_j}{\partial\lambda_p}\rangle .
\end{eqnarray}
The formal solution of the  time-dependent Schr$\ddot{o}$dinger equation gives the matrix Berry phase $\Phi_C=Pe^{ -\oint_C\sum_p A_p d\lambda_p}$,
where $P$ represents a path ordering.
Under a unitary transformation
$| \psi'_i \rangle =\sum_j U^*_{i j}| \psi_j \rangle$
the non-Abelian gauge structure emerges
\begin{eqnarray}
A'_k=UA_kU^\dagger+i U \frac{\partial U^\dagger}{\partial \lambda_k}.
\label{eq:vec_pot}
\end{eqnarray}

Here we propose that quantum dots in n-type semiconductors with spin-orbit interaction can have matrix Berry phases.
All the   discrete energy levels of  quantum dots possess  a double degeneracy 
because  the Rashba and/or Dresselhaus spin orbit coupling terms have time reversal symmetry.
In II-VI semiconductors
the Rashba term is expected to be larger than the
the Dresselhaus coupling. In III-V semiconductors, such as GaAs, the
opposite is true\cite{Rashba2}. In our work both the Rashba and Dresselhaus terms are included.
The adiabatic transformation
can be performed electrically by changing the confinement potentials of the quantum dot, see Fig.\ref{fig:setup}(d). 
We find that, although the spin orbit terms formally give  rise to a
non-Abelian structure for the  matrix vector potentials, 
double degeneracy does not necessarily lead to 
finite non-Abelian Berry phases.
We show  that when  the circular symmetry of two-dimensional harmonic potentials is broken  
matrix Berry phases can be  produced (such a distortion is shown in Fig.\ref{fig:setup}(b)).
We propose that the presence of a matrix Berry phase may be detected by measuring the  
the optical spectrum.

In Sec. II we describe the   Hamiltonian of the system  in detail, and in Sec. III we discuss 
a  model calculation of the matrix Berry phase. A possible experimental
detection of the the matrix Berry phase is suggested in Sec. IV.  Conclusions  are given in Sec. V.

\section{Hamiltonian}
An electric field $E$ is applied 
along the z-axis and electrons are confined in a triangular potential $V(z)$, see Fig.\ref{fig:setup}(a).
When the the width of the quantum well along the z-axis is sufficiently {\it small} we may include only the lowest
subband state along  the z-axis. We denote this  wavefunction by $f(z)$, see Fig.\ref{fig:setup}(a).
From the expectation value
$\langle f(z)|k_z^2|f(z)\rangle =0.8(2m^*eE/\hbar^2)^{2/3}$
we estimate the characteristic length scale along the z-axis: $R_z=1/\sqrt{0.8(2m^*eE/\hbar^2)^{2/3}}$.
The Hamiltonian in the absence of the spin orbit coupling is 
$H_K=-\frac{\hbar^2\nabla^2}{2m^*}+U(x,y)+V(z)$.
We take the two-dimensional potential to be $U(x,y)=\frac{1}{2}m^*\omega^2_x x^2+\frac{1}{2}m^*\omega^2_y y^2+
V_p(x,y)$, see Fig.\ref{fig:setup}(b).
The strengths of the harmonic potentials are denoted by $\omega_x$ and $\omega_y $.   They 
may be varied by {\it changing gate potentials} of the quantum dot system. 
The characteristic lengths scales along x- and y-axis are $R_{x,y}=\sqrt{\frac{\hbar}{m^*\omega_{x,y}}}$.
The potential  $V_p(x,y)=\epsilon' y$ represents a  distortion of the 
harmonic potentials and its strength $\epsilon'$ may be {\it varied electrically}. 

In a periodic crystal potential of a semiconductor 
the spin orbit interaction has two contributions.
The Rashba spin orbit term is 
\begin{eqnarray}
H_\mathrm{R}=c_\mathrm{R} \left( \sigma_x k_y -\sigma_y k_x \right).
\end{eqnarray}
Here $\sigma_{x,y}$ are Pauli spin matrices and $k_{x,y}$ are momentum operators ($k_x=\frac{1}{i}\frac{d}{dx}$
and similarly with $k_y$.).  
The constant $c_R$ {\it depends on the external electric field} $E$ applied along the z-axis.
The Dresselhaus spin orbit term is
\begin{eqnarray}
H_\mathrm{D}=c_\mathrm{D}\left( 
\left( \sigma_x k_x \left(k_y^2-k_z^2 \right) \right)
+\left( \sigma_y k_y \left(k_z^2-k_x^2 \right) \right)
\right).
\label{Dresselhaus}
\end{eqnarray}
There is another term of the form  $ \sigma_z \langle k_z\rangle \left(k_x^2-k_y^2 \right) $ in the  Dresselhaus spin orbit term
but it vanishes  since the expectation value
$\langle k_z\rangle =\langle f(z)|k_z|f(z)\rangle =0$ for the first subband  along  z-axis.
The  constant $c_D$  represents breaking of inversion symmetry by the crystal  in zinc blende structures.

The total Hamiltonian of an electron in  a semiconductor quantum dot
is
$H=H_\mathrm{K}+H_\mathrm{R}+H_\mathrm{D}$.
An  eigenstate of the Hamiltonian $H$ may be expanded as a linear combination of the eigenstates of $H_K$ 
\begin{eqnarray}
| \psi \rangle&=&\sum_{m n}c_{m n\uparrow }|m n \uparrow \rangle 
+\sum_{m n}c_{m n\downarrow }|m n \downarrow \rangle .
\end{eqnarray}
The expansion coefficients 
satisfy a matrix equation 
$\sum_{m'n'\sigma'}\langle mn\sigma|H|m'n'\sigma'\rangle c_{m'n'\sigma'}=Ec_{mn\sigma}$.
In the basis states $|m n \sigma \rangle $ the quantum number $m (n)$ and $\sigma$ label 
the harmonic oscillator levels along  the x-axis (y-axis) and
the component of electron spin.  The subband wavefunction $f(z)$ is suppressed in the notation $|m n \sigma \rangle$.

In the absence of the Zeeman term the total Hamiltonian is invariant under time reversal symmetry:
$\vec{k}\rightarrow -\vec{k}$ and $\vec{\sigma}\rightarrow -\vec{\sigma}$.
The time reversal operator is 
$K=-i \sigma_y C$,
where the operator $C$ stands for complex conjugation. The time reversed state of $ | \psi \rangle$ is
\begin{eqnarray}
| \overline{\psi} \rangle=K | \psi \rangle
=-\sum_{m n}c_{m n \downarrow}^* |m n \uparrow \rangle
+\sum_{m n} c_{m n\uparrow}^* |m n \downarrow \rangle.
\end{eqnarray}
Note that $K^2| \psi \rangle=-| \psi \rangle$.
These two states are degenerate and   orthonormal.  
We have suppressed the Bloch wavefunction of the conduction band in applying the time reversal
operator since it is unaffected by the operator $K$. Our wavefunctions are 
all effective mass wavefunctions  and only the  conduction band Bloch
wavefunction at $\vec{k}=0$ is relevant.

An  adiabatic change implemented 
by changing the energy parameters  $E_R$ and $E_p$ that characterize the Rashba term and the distortion potential $V_p(x,y)$:
\begin{eqnarray}
E_R=\frac{c_R}{\sqrt{2}R_y} \ \textrm{and}\ E_p=\epsilon R_y,
\label{parameters}
\end{eqnarray}
where
$\epsilon=\epsilon'/\sqrt{2}$.
The first parameter  $E_R$ 
depends on the electric field through the constant $c_R$.  
The typical value of  the energy scale associated with 
the Rashba constant  depends on the electric field applied along the z-axis and the semiconductor material: it
is of  order  $E_R=c_R/ R\sim 0.01-10meV$, where the   length
scale $R\sim 100 \AA$ is the lateral dimension of the quantum dot.
The second adiabatic parameter $E_p$ represents the strength of the distortion potential $\epsilon' y$:
the expectation value of the distortion potential is $E_p=\langle 0|\epsilon' y|1\rangle$.  Its 
magnitude is of order $1-10meV$, depending on the electric field applied along the y-axis and the
width of the triangular potential $V(z)$.

\section{Model calculation of non-Abelian Berry phase}    
\subsection{A truncated Hamiltonian matrix}

We work out a  model that can be solved analytically.
This model is simple but much can be learned from it.
Let us take $\omega_x=2\omega_y$ (Other values of $\omega_x$ can also be chosen). Then the lowest eigenenergy state
of $H_K$ is $|mn\rangle=|00\rangle$ with the energy $E_0=\frac{3}{2}\hbar\omega_y$ and 
the next lowest eigenenergy state  is $|01\rangle$ with the energy $E_1=\frac{5}{3}E_0$.
The typical value of the energy spacing between the quantum dot levels, $E_0$, is of  order $1-10 meV$.
Let us write down the eigenstates of the total Hamiltonian  as a  linear combination
of four basis states made out of these
states and spin degree of freedom:
\begin{eqnarray}
| \psi \rangle&=&
c_{0,0,\uparrow}|0,0, \uparrow \rangle+c_{0,1,\uparrow}|0,1, \uparrow \rangle  \nonumber\\
&+&c_{0,0,\downarrow}|0,0, \downarrow \rangle+c_{0,1,\downarrow}|0,1, \downarrow \rangle.
\label{eq:lin_comb1}
\end{eqnarray}
The  $4\times4$ truncated Hamiltonian matrix is
\begin{eqnarray}
\left(
\begin{array}{cccc}
E_0 & E_p & 0 & -i E_R - E_D \\
E_p & E_1 & i E_R +E_D & 0 \\
0 & -i E_R + E_D & E_0 & E_p \\
i E_R -E_D & 0 & E_p & E_1
\end{array}
\right),\nonumber\\
\label{matrix_Hamil}
\end{eqnarray}
or
\begin{eqnarray}
&&\frac{1}{2}(E_0+E_1)I
+\frac{1}{2}(E_0-E_1)\left( \begin{array}{cc} \sigma_z & 0 \\ 0 & \sigma_z \end{array} \right) \nonumber\\
&&+E_p \left( \begin{array}{cc} \sigma_x & 0 \\ 0 & \sigma_x \end{array} \right)
+E_R \left( \begin{array}{cc} 0 & \sigma_y \\ \sigma_y & 0 \end{array} \right)\nonumber\\
&&+E_D \left( \begin{array}{cc} 0 & -i \sigma_y \\ i \sigma_y & 0 \end{array} \right).
\end{eqnarray}
The matrix elements of the distortion potential are
$\langle m|y|n\rangle=\sqrt{\frac{\hbar}{2 m^* \omega_y}} (\sqrt{n+1} \delta_{m,n+1}+\sqrt{n} \delta_{m,n-1})$. 
Note that the distortion potential couples even and odd parity
states of the one-dimensional harmonic potential.  Other functional form of $V_p(x,y)$ may be also
used to generate the non-Abelian Berry phase as long as it couples even and odd parity
states.
The first term in the Dresselhaus spin orbit term, Eq. (\ref{Dresselhaus}), is zero since $\langle 0|k_x|0\rangle=0$. 
From the second term of the Dresselhaus term we get
$-ic_D\langle 0|k_y|1\rangle(\langle f(z)|k_z^2|f(z)\rangle-\langle 0|k_x^2|0\rangle)=-E_D$,
where the constant $E_D=f(E_R)-g(E_0)$ with
$f(E_R)=c_D/R_yR_z^2$
and
$g(E_0)=c_D /R_yR_x^2$.
Here we have used
the momentum matrix elements 
$\langle m|\hbar k_{x,y}|n\rangle=i \sqrt{\frac{m^* \hbar \omega_{x,y}}{2}} (\sqrt{n+1} \delta_{m,n+1}-\sqrt{n} \delta_{m,n-1})$.
The magnitude of the energy scale associated with the Dresselhaus term is of order  $E_D=c_D/ R^3$, and
it can be larger or smaller than the Rashba term, depending on the material\cite{Rashba2}.
The function $f(E_R)$ depends on  $E_R$ because
the  electric field $E$ enters through $R_z$. 
There is no simple way to calculate $c_R$ because it depends both on the electric field inside the 
semiconductor heterostructure and on the detailed boundary conditions at the interface.
For simplicity we take $f(E_R)=aE_R$  and $g(E_0)=bE_0$, where $a$ and $b$ are numerical constants.
More complicated functions, for example, $f(E_R)=E_R^2/E_0$ and $g(E_0)=E_0$, could be used, but our calculation indicates that 
the essential physics does not change.

\subsection{Finite matrix Berry phase }
Diagonalization of  this $4\times4$ Hamiltonian matrix
gives the eigenenergies
$\lambda=\frac{1}{3}(4E_0\pm \sqrt{E_0^2+9E_D^2+9E_p^2 +9E_R^2})$.
Let us choose the following  state
\begin{eqnarray}
|\psi\rangle=
\frac{1}{N}
\left(
\begin{array}{c}
3 E_p \\
E_0-\sqrt{E_0^2+9(E_D^2+E_p^2+E_R^2)} \\
3(E_D-i E_R) \\
0
\end{array}
\right),\nonumber\\
\label{eq:lin_comb2}
\end{eqnarray}
and its time reversal state  
\begin{eqnarray}
|\overline{\psi}\rangle=
\frac{1}{N^*}
\left(
\begin{array}{c}
-3(E_D+i E_R) \\
0 \\
3 E_p \\
E_0-\sqrt{E_0^2+9(E_D^2+E_p^2+E_R^2)}
\end{array}
\right).\nonumber\\
\end{eqnarray}
as the basis states in the lowest energy degenerate Hilbert space.

In the evaluation of the matrix $A_p$ we use 
$2\int \phi_k(r)^*\frac{\partial\phi_k(r)}{\partial \lambda_p}=\frac{\partial}{\partial \lambda_p}
\int|\phi_k(r)|^2dr=0$,
where  $k=mn\sigma$.  Note that
$\phi_k(r)$ is the two-dimensional harmonic oscillator wavefunction and that it is a real function. 
One can show that
$(A_p)_{i,j}=i \sum_k \alpha_k^*\frac{\partial \beta_k}{\partial\lambda_p}$,
where the pair of degenerate states are 
$| \psi_i \rangle=\sum_k \alpha_k|k\rangle$   and   $| \psi_j \rangle=\sum_k \beta_k|k\rangle$.
The orthonormalization  $\langle\psi_i|\psi_j\rangle=\delta_{ij}$    gives that the diagonal matrix elements
$(A_p)_{i,i}$ are real   and that the off-diagonal elements satisfy $(A_p)_{i,j}=(A_p)^*_{j,i}$.
The adiabatic parameters are  $\lambda_1=E_R$, $\lambda_2=E_p$.
We calculate the matrix Berry phase for
$E_D=E_R-E_0$. 
The matrix vector potentials with respect to $|\psi\rangle$   and $|\overline{\psi}\rangle$   have the following structures 
\begin{eqnarray}
A_{E_R}
&=&
\left(
\begin{array}{cc}
c_1 & a_1+i b_1 \\
a_1-i b_1 & -c_1
\end{array}
\right) \nonumber \\
&=&a_1 \sigma_x -b_1 \sigma_y +c_1 \sigma_z,  
\end{eqnarray}
and 
\begin{eqnarray}
A_{E_p}
&=&
\left(
\begin{array}{cc}
0 & a_2-i b_2 \\
a_2+i b_2 & 0
\end{array}
\right)\nonumber \\
&=&a_2 \sigma_x +b_2 \sigma_y. 
\end{eqnarray}
The functions  $a_i$, $b_i$, and $c_i$ depend on $E_R$ and $E_p$ and are real. 
Under an adiabatic time evolution a state in this degenerate Hilbert space changes as 
$|\psi'(t)\rangle=c_1(t) |\psi(t)\rangle + c_2(t) |\overline{\psi}(t)\rangle$.
Suppose initially  $c_1(0)=1$ and $c_2(0)=0$, i.e., $|\psi'(0)\rangle=|\psi\rangle$.
The cyclic adiabatic path with the period $T=2\pi/\omega$ is shown in Fig.\ref{fig:setup}(d) and is given by 
\begin{eqnarray}
&(&E_R(t),E_p(t))= \nonumber \\
&(&E_{R,c}+\Delta E_R \cos(\omega t),E_{p,c}+\Delta E_p \sin(\omega t)).
\label{closed_loop}
\end{eqnarray}
The frequency $\omega$ can be taken to be a fraction of $E_0$.
We solve Eq.(\ref{eq:time_Schrod})   numerically  and find that $c_2(T)$ is non-zero.
For the parameters $E_{R,c}= 2 E_0$, $E_{p,c}=E_0$, $\Delta E_R= 1.9 E_0$, $\Delta E_p= 0.9 E_0 $, and $\omega=E_0 /10$
we find $c_1(T)=0.8884-i 0.0897$  and $c_2(T)=-0.4429 +i 0.0874$.

\subsection{Absence of matrix  Berry phase}
The matrix Berry phase is absent when the distortion potential $V_p(x,y)$ is zero.
When only Rashba and/or Dresselhaus terms are present 
in  Eq.(\ref{matrix_Hamil})
the adiabatic transformation can be performed
 by varying the parameters $\lambda_1=E_R$  and $\lambda_2=E_0$.
In this case  {\it an orthonormal basis set exists} in the degenerate Hilbert space such that
the matrix vector potentials are diagonal and the non-Abelian 
Berry phase is zero. This degenerate basis set,  $|\psi_1\rangle$ and $|\psi_2\rangle$, has the property that
for each $k$  either $c_k^{(1)}=0$
or  $c_k^{(2)}=0$, where $|\psi_i\rangle=\sum_{k=1}^{4}c_k^i|k\rangle$
(see, for example, Eqs.(\ref{eq:finite_B_1}) and (\ref{eq:finite_B_2})).
The off-diagonal matrix elements of the vector potential $(A_p)_{1,2}
=i \langle\psi_1| \frac{d\psi_2}{d\lambda_p}\rangle=i \sum_{k=1}^{4}c_k^{(1)*}
\frac{dc^{(2)}_{k}}{d\lambda_p}=0$
because either $c_k^{(1)}=0$
or  $c_k^{(2)}=0$.
Therefore the matrix vector potentials are diagonal.
In this case  the matrix Berry phase will be absent.
Let us construct  explicitly  $| \psi_1 \rangle$ and $| \psi_2 \rangle$  when only the  Rashba term is present, i.e., when $E_D=E_p=0$.
They are given  by
\begin{eqnarray}
| \psi_1 \rangle&=&\frac{1}{\sqrt{2} \sqrt{E_0^2+9 E_R^2+E_0 \sqrt{E_0^2+ 9 E_R^2}}}  \nonumber\\
&\times&
\left(
\begin{array}{c}
i(E_0+\sqrt{E_0^2+9 E_R^2})\\
0 \\
0 \\
3 E_R
\end{array}
\right),\nonumber\\
\label{eq:finite_B_1}
\end{eqnarray}
and
\begin{eqnarray}
| \psi_2 \rangle&=&\frac{1}{\sqrt{9 E_R^2+\left(E_0 -\sqrt{E_0^2+ 9 E_R^2}\right)^2}} \nonumber\\
&\times&
\left(
\begin{array}{c}
0 \\
i (E_0 -\sqrt{E_0^2+9 E_R^2}) \\
3 E_R \\
0
\end{array}
\right).\nonumber\\
\label{eq:finite_B_2}
\end{eqnarray}
With these new  eigenstates it is possible to show that the non-Abelian gauge potentials 
are not only diagonal but also zero: $A_{E_0}=0$ and $A_{E_R}=0$.

\subsection{Non-Abelian gauge structure}
Let us also test the non-Abelian gauge structure given by   Eq. (\ref{eq:vec_pot}).
We make a unitary transformation from ($| \psi_1 \rangle$, $| \psi_2 \rangle$), given in Eqs.(\ref{eq:finite_B_1})
and (\ref{eq:finite_B_2}),
to a  pair of time reversed degenerate eigenstates 
$(| \psi'\rangle,| \overline{\psi}'\rangle)$,
\begin{eqnarray}
| \psi' \rangle=\frac{3}{2 \sqrt{E_0^2+9E_R^2}}
\left(
\begin{array}{c}
E_R \\
E_R \\
\frac{1}{3} i \left(E_0+\sqrt{E_0^2+9E_R^2} \right) \\
-\frac{1}{3} i \left(-E_0+\sqrt{E_0^2+9E_R^2} \right)
\end{array}
\right),\nonumber\\
\end{eqnarray}
and 
\begin{eqnarray}
| \overline{\psi}' \rangle=\frac{3}{2 \sqrt{E_0^2+9E_R^2}}
\left(
\begin{array}{c}
\frac{1}{3} i \left( E_0+\sqrt{E_0^2+9E_R^2} \right) \\
-\frac{1}{3} i \left( -E_0+\sqrt{E_0^2+9E_R^2} \right) \\
E_R \\
E_R \\
\end{array}
\right). \nonumber\\
\end{eqnarray}
Using these new basis states  one can show that    the non-Abelian gauge potentials are  
\begin{eqnarray}
A'_{E_R}
&=&\frac{1}{E_0^2+ 9 E_R^2}
\left(
\begin{array}{cc}
0 & 3E_0/2 \\
3E_0/2 & 0
\end{array}
\right) \nonumber \\
&=&\frac{3 E_0}{2(E_0^2+ 9 E_R^2)}\sigma_x,
\end{eqnarray}
and 
\begin{eqnarray}
A'_{E_0}
&=&\frac{1}{E_0^2+ 9 E_R^2}
\left(
\begin{array}{cc}
0 & -3E_R/2 \\
-3E_R/2 & 0
\end{array}
\right) \nonumber \\
&=&-\frac{3 E_R}{2(E_0^2+ 9 E_R^2)}\sigma_x.
\end{eqnarray}
Since the old vector potentials are $A_{E_R}=0$ and $A_{E_0}=0$ it  follows from  Eq. (\ref{eq:vec_pot}) 
that $A'_k=i U \frac{\partial U^\dagger}{\partial \lambda_k}$.
We have explicitly verified that this relation holds by computing the unitary matrix $U$.
We have also verified independently the absence of  the matrix Berry phase
by solving the time-dependent Schr$\ddot{o}$dinger Eq.(\ref{eq:time_Schrod}) 
with $A'_{E_R}$  and $A'_{E_0}$.  This provides a check on our numerical method of 
solving the time-dependent Schr$\ddot{o}$dinger equation.

\section{Detection of matrix Berry phase} 

After an adiabatic cycle the electron acquires a matrix Berry phase.
The presence of such a phase may be detected by measuring the strength of dipole optical transitions before
and after the cycle.  In this section we calculate the optical strengths using
the truncated Hamiltonian.  This calculation is not quantitatively accurate but
it will give us an estimate  of the   magnitude of the effect.

\begin{figure}[hbt]
\begin{center}
\includegraphics[width = 0.44 \textwidth]{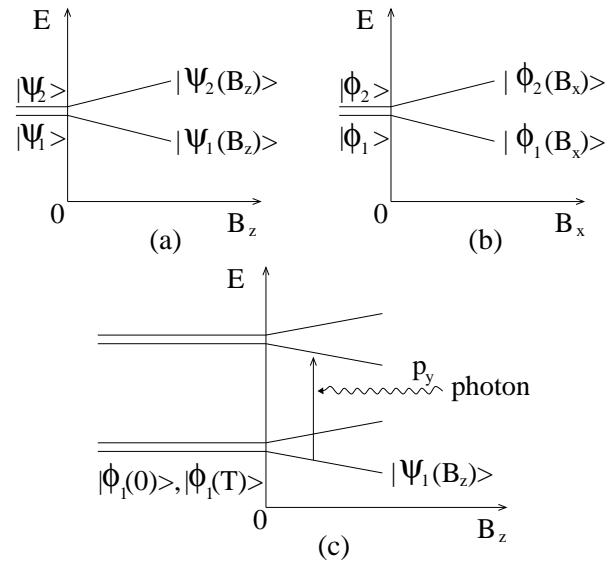}
\caption{(a)$|\psi_1\rangle$ and $|\psi_2\rangle$ are generated in the zero limit of $B_z$.
(b)$|\phi_1\rangle$ and $|\phi_2\rangle$ are generated in the zero limit of $B_x$.
(c)Dipole transition from the first to third lowest energy states in a finite  $B_z$.  The initial
state in this transition is   $|\phi_1(0) \rangle$ or  $|\phi_1(T) \rangle$.  These states represent, respectively,
the states before and after the adiabatic cycle.  The incident photon is polarized along the y-axis.
}
\label{fig:mag_field_limit}
\end{center}
\end{figure}

First we need to prepare physically some particular pair of  degenerate 
states.  In the presence of a magnetic field  
along the z-axis any degenerate pair of states at zero magnetic field  will be split into two
states. 
We  define  the lowest energy  pair of  degenerate eigenstates in the zero magnetic field limit of $B_z$ 
as $|\psi_1\rangle=\lim_{B_z\rightarrow 0}|\psi_1(B_z)\rangle$
and  $|\psi_2\rangle=\lim_{B_z\rightarrow 0} |\psi_2(B_z)\rangle$ , where $|\psi_1(B_z)\rangle$ and 
$|\psi_2(B_z)\rangle$ are the split lowest and next lowest energy states in a finite magnetic field, see Fig.(\ref{fig:mag_field_limit}a).
In a similar way we define degenerate eigenstates in the zero magnetic field limit of $B_x$:
$|\phi_1\rangle=\lim_{B_x\rightarrow 0}|\phi_1(B_x)\rangle$
and  $|\phi_2\rangle=\lim_{B_x\rightarrow 0} |\phi_2(B_x)\rangle$, see Fig.(\ref{fig:mag_field_limit}b).
The states $|\psi_{1,2}(B_z)\rangle$ and $|\phi_{1,2}(B_x)\rangle$ are calculated from the Hamiltonian by replacing
$\vec{k}$ with $\vec{k}+\frac{e}{c}\vec{A}$ and adding the Zeeman term ($\vec{A}$ is the vector potential and $e>0$).

Suppose that the electron is in the lowest energy state.  In order to measure the matrix Berry phase we  perform the following set of procedures:
\begin{enumerate}
\item  
We apply a  magnetic field along the x-axis and take the zero field limit.   The resulting state $|\phi_1 \rangle$
(Fig.(\ref{fig:mag_field_limit}b))
can be written as a linear combination of  $|\psi_1 \rangle$
and  $|\psi_2 \rangle$ (Fig.(\ref{fig:mag_field_limit}a)): $|\phi_1 \rangle$=$c_1(0)|\psi_1 \rangle+c_2(0)|\psi_2 \rangle$ with  
$(|c_1(0)|^2,|c_2(0)|^2)=(1/2,1/2)$.  

\item We apply  adiabatically a magnetic field along the z-axis.  
We increase it to a value $B_z$. Then the probability for  the electron to be  in the state $|\psi_1(B_z)\rangle$ 
is given by $|c_1(0)|^2$.

\item We then measure the intensity of the dipole transition  from the lowest energy state to the third lowest
energy state, see Fig.(\ref{fig:mag_field_limit}c).  This intensity is proportional to $|c_1(0)|^2$, i.e, proportional
to the probability that   $|\psi_1(B_z)\rangle$
is occupied.
The lowest and third lowest energy  states can be written as  a 
linear combination of the basis states $|mn\sigma \rangle$ : both states can be written in the form 
$ c_{0,0,\uparrow}|0,0, \uparrow \rangle+c_{0,1,\uparrow}|0,1, \uparrow \rangle
+c_{0,0,\downarrow}|0,0, \downarrow \rangle+c_{0,1,\downarrow}|0,1, \downarrow \rangle$.
When the photon is polarized along the y-axis  
only the basis states with different parities are coupled in   the dipole approximation,  for example,
$\langle 00\sigma|k_y|01\sigma\rangle=- \langle 01\sigma|k_y|00\sigma\rangle$ is non-zero.
\end{enumerate}
Again we assume that the electron is in the lowest energy state.  We  perform the second  set of procedures:
\begin{enumerate}
\item  We apply a  magnetic field along the x-axis and take the zero field limit.

\item We then perform an adiabatic cycle following a closed loop 
in the parameter space of $(E_R,E_p)$, given by Eq.(\ref{closed_loop}).
After
the  adiabatic cycle the electron will be in the state $|\phi_1(T) \rangle=c_1(T)|\psi_1 \rangle+c_2(T)|\psi_2 \rangle$ 
with $(|c_1(T)|^2,|c_2(T)|^2)=(0.559,0.441)$
(the parameters are $E_{R,c}=E_0$, $E_{p,c}=3E_0$, $\Delta E_R= 0.8E_0 $, $\Delta E_p= 2.5E_0$, and $\omega=0.2 E_0 $).

\item We apply adiabatically a magnetic field along the z-axis, see Fig.(\ref{fig:mag_field_limit}c).   Then the probability for  the electron 
to be in the first  lowest energy state, $|\psi_1(B_z)\rangle$, is $|c_1(T)|^2$
while  the probability that the electron
will be in the second lowest energy state, $|\psi_2(B_z)\rangle$, is $|c_2(T)|^2$.
Here the value of $B_z$ is the same as that of step 2 in the first set of procedures.

\item We then measure the intensity of the dipole transition  from the lowest energy state to the third lowest
energy state.  This intensity is proportional to $|c_1(T)|^2$,  i.e, proportional
to the probability that   $|\psi_1(B_z)\rangle$
is occupied.
\end{enumerate}

These two sets of measurements are repeated many times.   Then the ratio 
$\frac{|c_1(T)|^2 }{|c_1(0)|^2}=1.12$ gives the intensity ratio of the dipole transitions  in the first and second 
sets of procedures (the dipole matrix elements cancel each other).   It is the direct measure of 
the matrix Berry phase.   The analysis of the intensity of optical transitions in zero magnetic field is 
complicated due to the presence of the degeneracy in the final states of the transition. 
In a finite magnetic field this degeneracy is lifted.

\section{Discussions}

Each discrete eigenstate  of a semiconductor quantum dot with the Rashba and/or Dresselhaus spin orbit coupling terms
possesses  a double degeneracy due to time reversal symmetry.
We have investigated the matrix Berry phase of such a quantum dot 
in a simple truncated model Hamiltonian that can be solved analytically.
We have found that the double degeneracy  does not necessarily  lead to  a finite 
non-Abelian Berry phase.  
The addition of a parity breaking distortion potential to the Hamiltonian when both the  Rashba and Dresselhaus spin orbit coupling terms
are present  gives rise to
a finite matrix Berry phase.
We have proposed that this phase may be detected in the dipole  transitions
between the ground and first excited states in a magnetic field.

For an accurate modeling of possible experiments
the number of basis vectors in the truncated Hamiltonian matrix must be chosen 
sufficiently large.  This number  will be determined by the ratios  $E_0/E_{R,D,p}$.
Calculation of the matrix Berry phase in such a case requires a heavy numerical computation.
Accurate experimental determination of the functional dependence of the Rashba  constant on the electric field 
would be also valuable
in determining the magnitude of matrix Berry phases.
The quantum dot studied in this work contains  a single electron.
It may be worthwhile to investigate the effect of many body interactions.

\begin{acknowledgments}
This work was  supported by grant No. R01-2005-000-10352-0 from the Basic Research Program
of the Korea Science and Engineering
Foundation and by Quantum Functional Semiconductor Research Center (QSRC) at Dongguk University
of the Korea Science and Engineering
Foundation. 
\end{acknowledgments}

\end{document}